\magnification=\magstep1 \overfullrule=0pt
\parskip=6pt
\baselineskip=15pt
\headline={\ifnum\pageno>1 \hss \number\pageno\ \hss \else\hfill \fi}
\pageno=1
\nopagenumbers
\hbadness=1000000
\vbadness=1000000

\centerline{\bf ON THE CALCULATION OF GROUP CHARACTERS}
\medskip

\centerline{\bf M. Gungormez}
\centerline{Dept. Physics, Fac. Science, Istanbul Tech. Univ.}
\centerline{80626, Maslak, Istanbul, Turkey }
\centerline{e-mail: gungorm@itu.edu.tr}

\medskip

\centerline{\bf H. R. Karadayi}
\centerline{Dept. Physics, Fac. Science, Istanbul Tech. Univ.}
\centerline{80626, Maslak, Istanbul, Turkey }
\centerline{e-mail: karadayi@itu.edu.tr }
\centerline{and}
\centerline{Dept. Physics, Fac. Science, Istanbul Kultur University}
\centerline{34156, Atakoy, Istanbul, Turkey }

\medskip
\centerline{\bf Abstract}
\vskip5mm

It is known that characters of irreducible representations of finite
Lie algebras can be obtained by Weyl character formula including
Weyl group summations which make actual calculations almost
impossible except for a few Lie algebras of lower rank. By starting
from Weyl character formula, we show that these characters can be
re-expressed without referring to Weyl group summations. Some useful
technical points are given in detail in the instructive example of
$G_2$ Lie algebra.

\vskip5mm
\vskip5mm
\vskip5mm
\vskip5mm
\vskip5mm
\vskip5mm

\eject

\vskip 3mm
\noindent {\bf{I.\ INTRODUCTION}}
\vskip 3mm

It is known that the character formula of Weyl {\bf [1]} gives us a
direct way to calculate the character of irreducible representations
of finite Lie algebras. For this, let $G_r$ be a Lie algebra of rank
r, $W(G_r)$ its Weyl group, $\alpha_i$'s and $\lambda_i$'s be,
respectively, its simple roots and fundamental dominant weights. The
notation here and in the following sections will be as in our
previous work {\bf [2]}. For further reading, we refer to the
beautiful book of Humphreys {\bf [3]}. A dominant weight $\Lambda^+$
is expressed in the form
$$ \Lambda^+ = \sum_{i=1}^r s_i \lambda_i  \eqno(I.1) $$
where $s_i$'s are some positive integers (including zero). An
irreducible representation $V(\Lambda^+)$ can then be attributed to
$\Lambda^+$ . The character $Ch(\Lambda^+)$ of $V(\Lambda^+)$ is
defined by
$$ Ch(\Lambda^+) \equiv \sum_{\lambda^+} \ \sum_{\mu \in W(\lambda^+)}
m_{\Lambda^+}(\mu) \ e^\mu \eqno(I.2) $$ where $m_{\Lambda^+}(\mu)$\
's are multiplicities which count the number of times a weight $\mu$
is repeated for $V(\Lambda^+)$. The first sum here is over
$\Lambda^+$ and all of its sub-dominant weights $\lambda^+$'s while
the second sum is over the elements of their Weyl orbits
$W(\lambda^+)$'s. Formal exponentials are taken just as in the book
of Kac {\bf [4]} and in (I.2) we extend the concept for any weight
$\mu$, in the form $e^\mu$. Note here that multiplicities are
invariant under Weyl group actions and hence it is sufficient to
determine only $m_{\Lambda^+}(\lambda^+)$ for the whole Weyl orbit
$W(\lambda^+)$.

An equivalent form of (I.2) can be given by
$$ Ch(\Lambda^+) = { A(\rho+\Lambda^+) \over A(\rho) }  \eqno(I.3) $$
where $\rho$ is the Weyl vector of $G_r$. (I.3) is the celebrated
{\bf Weyl Character Formula} which gives us the possibility to calculate
characters in the most direct and efficient way. The central objects here
are $A(\rho+\Lambda^+)$'s which include a sum over the whole Weyl group:
$$ A(\rho+\Lambda^+) \equiv \sum_{\sigma \in W(G_r)} \ \epsilon(\sigma) \
e^{\sigma(\rho+\Lambda^+)}   \ \   .  \eqno(I.4) $$
In (I.4), $\sigma$ denotes an element of Weyl group,i.e. a Weyl reflection,
and $\epsilon(\sigma)$ is the corresponding {\bf signature} with values
either +1 or -1.

\eject

The structure of Weyl groups is completely known for finite Lie algebras
in principle. In practice, however, the problem is not so trivial, especially
for Lie algebras of some higher rank. The order of $E_8$ Weyl group is, for
instance, 696729600 and any application of Weyl character formula needs for
$E_8$ an explicit calculation of a sum over 696729600 Weyl reflections.
Our main point here is to overcome this difficulty in an essential manner.

For an actual application of (I.3), an important notice is the
{\bf specialization} of formal exponentials $e^\mu$ as is called in the book
of Kac {\bf [4]}. In its most general form, we consider here the specialization
$$ e^{\alpha_i} \equiv u_i \  ,  \ i=1,2, \dots, r  .  \eqno(I.5) $$
which allows us to obtain $A(\rho)$ in the form of
$$ A(\rho) = P(u_1,u_2, \dots, u_r)  \eqno(I.6) $$
where $P(u_1,u_2, \dots, u_r)$ is a polynomial in indeterminates
$u_i$'s. We also have
$$ A(\rho+\Lambda^+) = P(u_1,u_2, \dots, u_r;s_1,s_2, \dots ,s_r) \eqno(I.7) $$
where $P(u_1,u_2, \dots, u_r;s_1,s_2, \dots, s_r)$ is another
polynomial of indeterminates $u_i$'s and also parameters $s_i$'s
defined in (I.1). The Weyl formula (I.3) then says that polynomial
(I.7) always factorizes on polynomial (I.6) leaving us with another
polynomial \break $R(u_1,u_2, \dots, u_r;s_1,s_2, \dots, s_r)$ which
is nothing but the character polynomial of $V(\Lambda^+)$.

The specialization (I.5) will always be normalized in such a way
that (I.3) gives us the {\bf Weyl dimension formula} {\bf [5]}, in
the limit $u_i = 1$ for all $i=1,2, \dots, r$. One also expects that
$$ P(u_1,u_2, \dots, u_r;0,0, \dots, 0) \equiv P(u_1,u_2, \dots, u_r) \ .
\eqno(I.8) $$

\vskip 3mm
\noindent {\bf{II.\ RECREATING $A(\rho+\Lambda^+)$ FROM $A(\rho)$ }}
\vskip 3mm

In this section, without any reference to Weyl groups, we give a way
to calculate polynomial (I.7) directly from polynomial (I.6). For this,
we first give the following explicit expression for polynomial (I.6):
$$ A(\rho) = {
\prod_{\alpha \in \Phi^+} \ ( e^\alpha - 1 ) \over
\prod_{i=1}^r \ (e^{\alpha_i})^{k_i}     }   \eqno(II.1)  $$
where
$$ k_i \equiv {1 \over 2} \ (\alpha_i,\alpha_i) \ (\lambda_i,\rho)
\eqno(II.2) $$
and $\Phi^+$ is the positive root system of $G_r$. Exponents $k_i$'s are due
to the fact that the monomial of maximal order is
$$ \prod_{i=1}^r \ (e^{\alpha_i})^{2 k_i}  $$
in the product $\prod_{\alpha \in \Phi^+} \ ( e^\alpha - 1 )$.
All the scalar products like $(\lambda_i,\alpha_j)$ or $(\alpha_i,\alpha_i)$
are the symmetrical ones and they are known to be defined via Cartan matrix of
a Lie algebra. The crucial point, however, is to see that (II.1) is equivalent to
$$ A(\rho) = \prod_{A=1}^{\vert W(G_r) \vert} \ \epsilon_A \ \
(e^{\alpha_i})^{\xi_i^0(A)}    \eqno(II.3)  $$ where $\vert W(G_r)
\vert$ is the order of Weyl group $W(G_r)$ and as is emphasized in
\break section I, $\epsilon_A$'s are signatures with values
$\epsilon_A = \mp 1$. Note here that, by expanding the product
$\prod_{\alpha \in \Phi^+} \ ( e^\alpha - 1 )$, the precise values
of signatures can be determined uniquely.

To get an explicit expression for exponents $\xi_i^0(A)$ in (II.3),
let us define ${\bf \it R^+}$ is composed out of elements of the
form
$$ \beta^+ \equiv \sum_{i=1}^r n_i \ \alpha_i   \eqno(II.4) $$
where $n_i$'s are some positive integers including zero. ${\bf \it R^+}$ is a subset
of the Positive Root Lattice of $G_r$. The main emphasis here is on some special roots
$$ \gamma_i(I_i) \in {\bf \it R^+} $$
which are defined by following conditions
$$ (\lambda_i - \gamma_i(I_i),\lambda_j - \gamma_j(I_j)) = (\lambda_i,\lambda_j) \ \  ,
\ \  i,j = 1,2, \dots, r .   \eqno(II.5)  $$
Note here that $(\lambda_i,\lambda_j)$'s are defined by inverse Cartan matrix.
For the range of indices $ I_j$'s, we assume that they take values from the set
$ \{1,2, \dots, \vert I_j \vert \} $, that is
$$ I_j \in \{1,2, \dots, \vert I_j \vert \} \ \ , \ \ j = 1,2, \dots, r. $$

\noindent We also define the sets

$$ \Gamma(A) \equiv \{ \gamma_1(I_1(A)),\gamma_2(I_2(A)), \dots, \gamma_r(I_r(A)) \} \ \ ,
\ \ A=1,2, \dots, D  \eqno(II.6)  $$

\noindent are chosen by conditions (II.5) where $I_j(A) \in \{1,2,
\dots, \vert I_j \vert \} $ and D is the maximal number of these
sets.

Following two statements are then valid:

\noindent (1) \ \ ${\bf  D = \vert W(G_r) \vert  }$

\noindent (2) \ \ ${\bf \vert I_j \vert  = \vert W(\lambda_j) \vert  }$

\noindent where $ \vert W(\lambda_j) \vert $ is the order, i.e.
the number of elements, of the Weyl orbit $W(\lambda_j)$. As is known,
a Weyl orbit is stable under Weyl reflections and hence all its elements
have the same length. It is interesting to note however that the lengths
of any two elements $\gamma_i(j_1)$ and $\gamma_i(j_2)$ could, in general,
be different while the statement (2) is, still, valid.

The exponents in (II.3) can now be defined by

$$ \xi_i^0(A) \equiv {1 \over 2} \ (\alpha_i,\alpha_i) \
(\lambda_i - \gamma_i(A),\rho)  \ \ . \eqno(II.7) $$

\noindent We then also state that the natural extension of (II.3) is
as in the following:

$$ A(\rho+\Lambda^+) = \prod_{A=1}^{\vert W(G_r) \vert} \ \epsilon_A \ \
(e^{\alpha_i})^{\xi_i(A)}    \eqno(II.8)  $$

\noindent for which the exponents are

$$ \xi_i(A) \equiv {1 \over 2} \ (\alpha_i,\alpha_i) \
(\lambda_i - \gamma_i(A),\rho+\Lambda^+)  \ \ . \eqno(II.9) $$

\noindent as a natural extension of (II.7). Signatures have the same
values in both expressions, (II.3) and (II.8). This reduces the
problem of explicit calculation of character polynomial
$Ch(\Lambda^+)$ to the problem of finding solutions to conditions
(II.5). It is clear that this is more manageable than that of using
Weyl character formula directly.

\vskip 3mm
\noindent {$\bf{III.\ AN \ EXAMPLE : G_2 }$}
\vskip 3mm

It is known that $G_2$ is characterized by two different root lengths
$$ (\alpha_1,\alpha_1) = 6 \ \ \ , \ \ \  (\alpha_2,\alpha_2) = 2
\eqno(III.1) $$

\noindent and also

$$ (\alpha_1,\alpha_2) = -3  \ \ .     \eqno(III.2) $$

\noindent Its positive root system is thus

$$ \Phi^+ = \{ \ \alpha_1 \ , \ \alpha_2 \ , \ \alpha_1+\alpha_2 \ , \
\alpha_1+2 \ \alpha_2 \ , \ \alpha_1+3 \ \alpha_2 \ , \
2 \ \alpha_1+3  \ \alpha_2 \ \}  \eqno(III.3)  $$

\noindent Beyond $\Phi^+$, first few elements of ${\bf \it R^+}$ are
determined by

$$ \eqalign{&
\ \ 2 \ \alpha_2 \ \                  \  ,
\ \ 2 \ \alpha_1 + 2 \ \alpha_2 \     \  ,
\ \ 2 \ \alpha_1 \ + \ 4 \ \alpha_2 \ \  ,
\ \ \alpha_1 \ + \ 4 \alpha_2 \       \  ,      \cr
&\ \ 2 \ \alpha_1 \ + \ \alpha_2 \    \  ,
\ \ 2 \ \alpha_1 \ + \ 5 \ \alpha_2 \ \  ,
\ \ 3 \ \alpha_1 \ + \ 4 \ \alpha_2 \ \  ,
\ \ 3 \ \alpha_1 \ + \ 5 \ \alpha_2 \ \  ,
\ \ 3 \ \alpha_2 \                    \  ,      \cr
&\ \ 3 \ \alpha_1 \ + \ 3 \ \alpha_2 \ \ ,
\ \ 3 \ \alpha_1 \ + \ 6 \ \alpha_2 \ \  ,
\ \ 2 \ \alpha_1 \                    \  ,
\ \ 2 \ \alpha_1 \ + \ 6 \ \alpha_2 \ \  ,
\ \ 4 \ \alpha_1 \ + \ 6 \ \alpha_2 \                 }    $$

\noindent as will be seen from the definition (II.4). This will be
sufficient to obtain following elements which fulfill conditions
(II.5):

$$ \eqalign{
& \gamma_1(1) = 0 \ , \cr
&\gamma_1(2) = \alpha_1 \ , \cr
&\gamma_1(3) = \alpha_1 \ + \ 3 \ \alpha_2 \ , \cr
&\gamma_1(4) = 3 \ \alpha_1 \ + \ 3 \ \alpha_2 \ , \cr
&\gamma_1(5) = 3 \ \alpha_1 \ + \ 6 \ \alpha_2 \ , \cr
&\gamma_1(6) = 4 \ \alpha_1 \ + \ 6 \ \alpha_2 \  } $$
and
$$ \eqalign{
& \gamma_2(1) = 0 \ , \cr
&\gamma_2(2) = \alpha_2 \ , \cr
&\gamma_2(3) = \alpha_1 \ + \ \alpha_2 \ , \cr
&\gamma_2(4) = \alpha_1 \ + \ 3 \ \alpha_2 \ , \cr
&\gamma_2(5) = 2 \ \alpha_1 \ + \ 3 \ \alpha_2 \ , \cr
&\gamma_2(6) = 2 \ \alpha_1 \ + \ 4 \ \alpha_2         } $$

\noindent from which we obtain,as in (II.6), following 12 sets of
2-elements:

$$ \eqalign{
\Gamma_1 & = \{ \gamma_1(1) , \gamma_2(1) \} \ \ , \ \ \epsilon_1 = +1     \cr
\Gamma_2 & = \{ \gamma_1(2) , \gamma_2(3) \} \ \ , \ \ \epsilon_2 = +1     \cr
\Gamma_3 & = \{ \gamma_1(3) , \gamma_2(2) \} \ \ , \ \ \epsilon_3 = +1     \cr
\Gamma_4 & = \{ \gamma_1(4) , \gamma_2(5) \} \ \ , \ \ \epsilon_4 = +1     \cr
\Gamma_5 & = \{ \gamma_1(5) , \gamma_2(4) \} \ \ , \ \ \epsilon_5 = +1     \cr
\Gamma_6 & = \{ \gamma_1(6) , \gamma_2(6) \} \ \ , \ \ \epsilon_6 = +1     \cr
\Gamma_7 & = \{ \gamma_1(1) , \gamma_2(2) \} \ \ , \ \ \epsilon_7 = -1     \cr
\Gamma_8 & = \{ \gamma_1(2) , \gamma_2(1) \} \ \ , \ \ \epsilon_8 = -1     \cr
\Gamma_9 & = \{ \gamma_1(3) , \gamma_2(4) \} \ \ , \ \ \epsilon_9 = -1     \cr
\Gamma_{10} & = \{ \gamma_1(4) , \gamma_2(3) \} \ \ , \ \ \epsilon_{10} = -1  \cr
\Gamma_{11} & = \{ \gamma_1(5) , \gamma_2(6) \} \ \ , \ \ \epsilon_{11} = -1  \cr
\Gamma_{12} & = \{ \gamma_1(6) , \gamma_2(5) \} \ \ , \ \ \epsilon_{12} = -1   }
\eqno(III.4)  $$

\noindent Note here that $\vert W(G_2) \vert = 12$, $\vert
W(\lambda_1) \vert = 6$ and $\vert W(\lambda_2) \vert = 6$ show, for
$G_2$, the validity of our two statements mentioned above. The
signatures are given on the right-hand side of (III.4). They can be
easily found to be true by comparing two expressions (II.1) and
(II.3) for the case in hand. As in (I.5), let us choose the most
general specialization
$$ e^{\alpha_1} = x \ \ , \ \ e^{\alpha_2} = y  \ \ . $$
Then, in view of (I.8), character polynomial
$$ R(x,y,s_1,s_2) \equiv { P(x,y,s_1,s_2) \over P(x,y,0,0) }  \eqno(III.5) $$
for an irreducible $G_2$ representation originated from a dominant weight \
$\Lambda^+ = s_1  \lambda_1 \ + \ s_2  \lambda_2$ \  will be expressed in
its following {\bf ultimate} form:
$$ \eqalign{
P(x,y,s_1,s_2) =
& + x^{ 3 + 2 \ s_1 + s_2} \ y^{ 5 + 3 \ s_1 + 2 \ s_2} \
  + x^{-3 - 2 \ s_1 - s_2} \ y^{-5 - 3 \ s_1 - 2 \ s_2}     \cr
& + x^{-2 - s_1 - s_2} \ y^{-1 - s_2}
  + x^{ 2 + s_1 + s_2} \ y^{ 1 + s_2}                       \cr
& + x^{-1 - s_1} \ y^{-4 - 3 \ s_1 - s_2}
  + x^{ 1 + s_1} \ y^{ 4 + 3 \ s_1 + s_2}                   \cr
& - x^{ 1 + s_1} \ y^{-1 - s_2}
  - x^{-1 - s_1} \ y^{ 1 + s_2}                             \cr
& - x^{-2 - s_1 - s_2} \ y^{-5 - 3 \ s_1 - 2 \ s_2}
  - x^{ 2 + s_1 + s_2} \ y^{ 5 + 3 \ s_1 + 2 \ s_2}         \cr
& - x^{-3 - 2 \ s_1 - s_2} \ y^{-4 - 3 \ s_1 - s_2}
  - x^{ 3 + 2 \ s_1 + s_2} \ y^{4 + 3 \ s_1 + s_2}             } \eqno(III.6)  $$

In the following we will give some examples to illustrate (III.5);
$$  \eqalign{
&Ch(\lambda_1)=R(x,y,1,0)={1 \over x^2 \ y^3} \ (1 + x + x \ y + x \
y^2 + x^2 \ y^2 + x \ y^3 + 2 \ x^2 \ y^3 + x^3 \ y^3 \cr & \ \ \ \
\ \ \ \ \ \ \ \ \ \ \ \ \ \ \ \ \ \ \ \ \ \ \ \ \ \ \ \ \ \ \ \ \ \
\ \ \ \ \ \ + x^2 \ y^4 + x^3 \ y^4 + x^3 \ y^5 + x^3 \ y^6 + x^4 \
y^6) \cr &Ch(\lambda_2)=R(x,y,0,1)={1 \over x \ y^2} \ (1 + y + x \
y + x \ y^2 + x \ y^3 + x^2 \ y^3 + x^2 \ y^4)    \cr
&Ch(\lambda_1+\lambda_2)=R(x,y,1,1)={1 \over x^3 \ y^5} \ (1 + x) \
(1 + y) \ (1 + x \ y) \ (1 + x \ y^2) \ (1 + \cr & \ \ \ \ \ \ \ \ \
\ \ \ \ \ \ \ \ \ \ \ \ \ \ \ \ \ \ \ \ \ \ \ \ \ \ \ \ \ \ \ \ \ \
\ \ \ \ \ \ \ \ x \ y^3) \ (1 + x^2 \ y^3)  \cr &Ch(2 \
\lambda_2)=R(x,y,0,2)={1 \over x^2 \ y^4} \ (1 + y + y^2) \ (1 + x \
y + x^2 \ y^2) \ (1 + x \ y^2 + x^2 \ y^4) } $$ where
$$  P(x,y,0,0)={1 \over x^3 \ y^5} (1 - x) \ (1 - y) \ (1 - x \ y) \ (1 - x \ y^2) \
(1 -x \ y^3) \ (1 - x^2 \ y^3)  $$

A useful application of (III.5) is in the calculation of tensor
coupling coefficients {\bf [6]}. We note, for instance, that the
coefficients in decomposition

$$ \eqalign{V(\lambda_1) \otimes V(\lambda_1 + \lambda_2)&=
V(2 \ \lambda_1 + \lambda_2) \oplus V(\lambda_1 +2 \ \lambda_2)
\oplus V(4 \ \lambda_2) \oplus V(3 \ \lambda_2) \cr &\oplus 2 \
V(\lambda_1 + \lambda_2) \oplus V(2 \ \lambda_2) \oplus V(\lambda_2)
} $$ \noindent can be calculated easily by
$$ \eqalign{R(x,y,1,0) \otimes R(x,y,1,1)&=
R(x,y,2,1) \oplus R(x,y,1,2) \oplus R(x,y,0,4) \oplus R(x,y,0,3) \cr
&\oplus 2 \ R(x,y,1,1) \oplus R(x,y,0,2) \oplus R(x,y,0,1) }$$
\noindent where the terms above are already known by (III.5).

\vskip 3mm
\noindent {\bf{CONCLUSION}}
\vskip 3mm

Although application of our method is given here in case of $G_2$
Lie algebra, there are no serious difficulty to do the same for
classical chains $A_r \ , \ B_r \ , \ C_r$ \ and \ $D_r$ of finite
Lie algebras. Our results for exceptional Lie algebras $F_4$ and
$E_6$ will be given soon in a subsequent publication. In case of
$E_7$ and $E_8$ Lie algebras, one must however emphasize that rather
than the most general specialization (I.5) of formal exponentials we
need in practice more simple specializations though there are some
other methods {\bf [7]} which work equally well for $E_7$ and $E_8$.
For Affine Lie Algebras, a similar study of Weyl-Kac character
formula will also be interesting.

\hfill\eject

\bigskip
\noindent{\bf {REFERENCES}}
\vskip5mm

\item{[1]} H.Weyl, The Classical Groups, N.J. Princeton Univ. Press (1946)

\item{[2]} H.R.Karadayi and M.Gungormez, J.Phys.A:Math.Gen. 32 (1999) 1701-1707

\item{[3]} J.E.Humphreys, Introduction to Lie Algebras and Representation Theory, \hfill\break
N.Y., Springer-Verlag (1972)

\item{[4]} V.G.Kac, Infinite Dimensional Lie Algebras, N.Y., Cambridge Univ. Press (1990)

\item{[5]} Weyl dimension formula cited in p.139 of ref [3]

\item{[6]} Steinberg formula cited in p.140 of ref [3]

\item{[7]} H.R.Karadayi and M.Gungormez, Summing over the Weyl Groups of $E_7$ and $E_8$, \hfill\break
math-ph/9812014

\end